# ZnO nanoforms: The state-of-the art of synthetic strategies

**Basavaraj S. Devaramani[1,#], Ramaswamy Y.S.[2],**
**Babu A. Manjasetty[3] and Gopalakrishna T.R. Nair [4,*]**

[1]M.Tech. Student (Electrical and Electronics), Research & Industry Incubation Center (RIIC), Dayananda Sagar Institutions(DSI), Bangalore. email: basavaraj@jncasr.ac.in
[#] Honeywell Technologies, Bangalore.

[2]Research Professor, Head, Nano Technology Initiative, RIIC, DSI, Bangalore.
email: euronicnano@gmail.com

[3]Research Professor, Proteomics and Bioinformatics Platform, RIIC, DSI, Bangalore.
email: babu.manjasetty@gmail.com

[4]Research Professor, Director, Software engineering Group, RIIC, DSI, Bangalore.
email: trgnair@yahoo.com

Dr. T. R. Gopalakrishna Nair, Ph.D
Director,
Research & Industry Incubation Center
Dayananda Sagar Institutions Shavige
Mallesha Hills Kumaraswamy Layout
Bangalore, India
email: trgnair@yahoo.com

## ABSTRACT

Nanotechnology has already made indelible dents in the fields of communication, computation, electronics, photonics, gas sensors, diagnostic technologies, catalysis, drug delivery systems and imaging. In particular, ZnO nanostructures have attracted a lot of research interest due to their unique structure- and size-dependent electrical, optical, and mechanical properties. ZnO nanosystems are abundant with respect to a wide variety of morphologies whose syntheses are critically dependent on several experimental parameters. Cursory presentation of the immense potential of these tiny systems and the state-of-the-art of the synthetic methodologies along with our initial research on ZnO nanorods has been presented. ZnO nanorods in the size range 50-150nm have been synthesized by solid-state route by heating zinc acetate with caustic soda under specific growth conditions. X-ray diffraction pattern unequivocally establishes a wurtzite structure. Scanning electron microscope images of ZnO nanorods display a random distribution of non-uniform size ranges. The manipulation of ZnO nanosystems by fine tuning the critical band gap of this unique semiconductor by doping with divalent transition metal ions leading to novel nanosystems to enhance the performance of the electronic devices is our extended area of research.

# ZnO nanoforms: The state-of-the art of synthetic strategies

## 1. Introduction

Nanotechnology is a powerful approach to integrate technologies from Physics, Chemistry, Engineering, Biology and Medicine (Figure 1). In recent years, it has led to a profound paradigm shift and is being classified as one of the most important areas of impending technology.

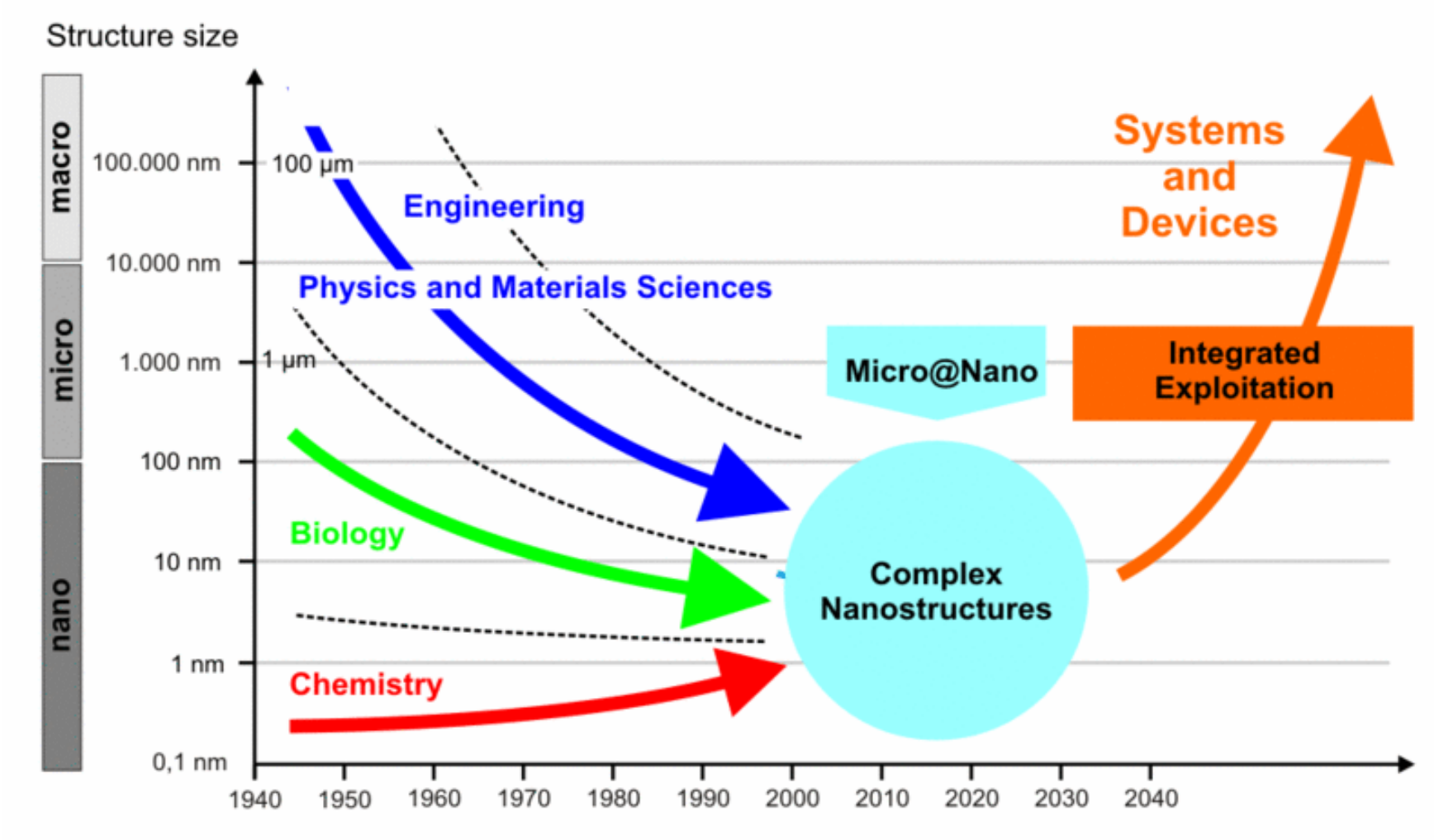

Figure 1. Graph represents a profile of the converging technologies with a plot of the time line versus particle size (http://www.nano.boku.ac.at/).

The research on nanostructures has rapidly expanded because of their unique and novel applications [1] in optics, optoelectronics, catalysis, biological sciences and piezoelectricity. Particularly, the ZnO nanosystems have exciting applications owing to their polymorphological structures [2]. This wide-band gap (energy of 3.37eV) semiconductor enables huge potential for electronic and optical applications [3]. It has unique piezoelectric properties that are very essential for fabricating devices or to enhance the performance of electromechanical devices [4]. It is a biocompatible material suitable for medical and biological applications [5]. ZnO nanosystems, such as wires, belts, needles and films can be easily formed by either chemical or physical approaches that are primarily temperature dependent. ZnO has polar surfaces that help in the formation of a wide range of nanostructures such as rings, springs, bows and helices [6]. An overview of the state-of-the-art of the synthetic methodologies along with our initial research results is reported.

## 2. Nano in preference to the bulk

Some man-made and nature-made devices and tools on nanometric scale are shown in figure 2. The length scale of the nanosystems is very crucial for various reasons. Firstly, confining electrons to a small space produces a nanostructure with novel properties such as ultra-strength, superior-elasticity, novel chemistry, unusual electrical and optical behavior that are completely size-dependent with high *surface-to-volume* ratio. The size of the nanomaterial becomes a property-tuning parameter especially in electronic devices.

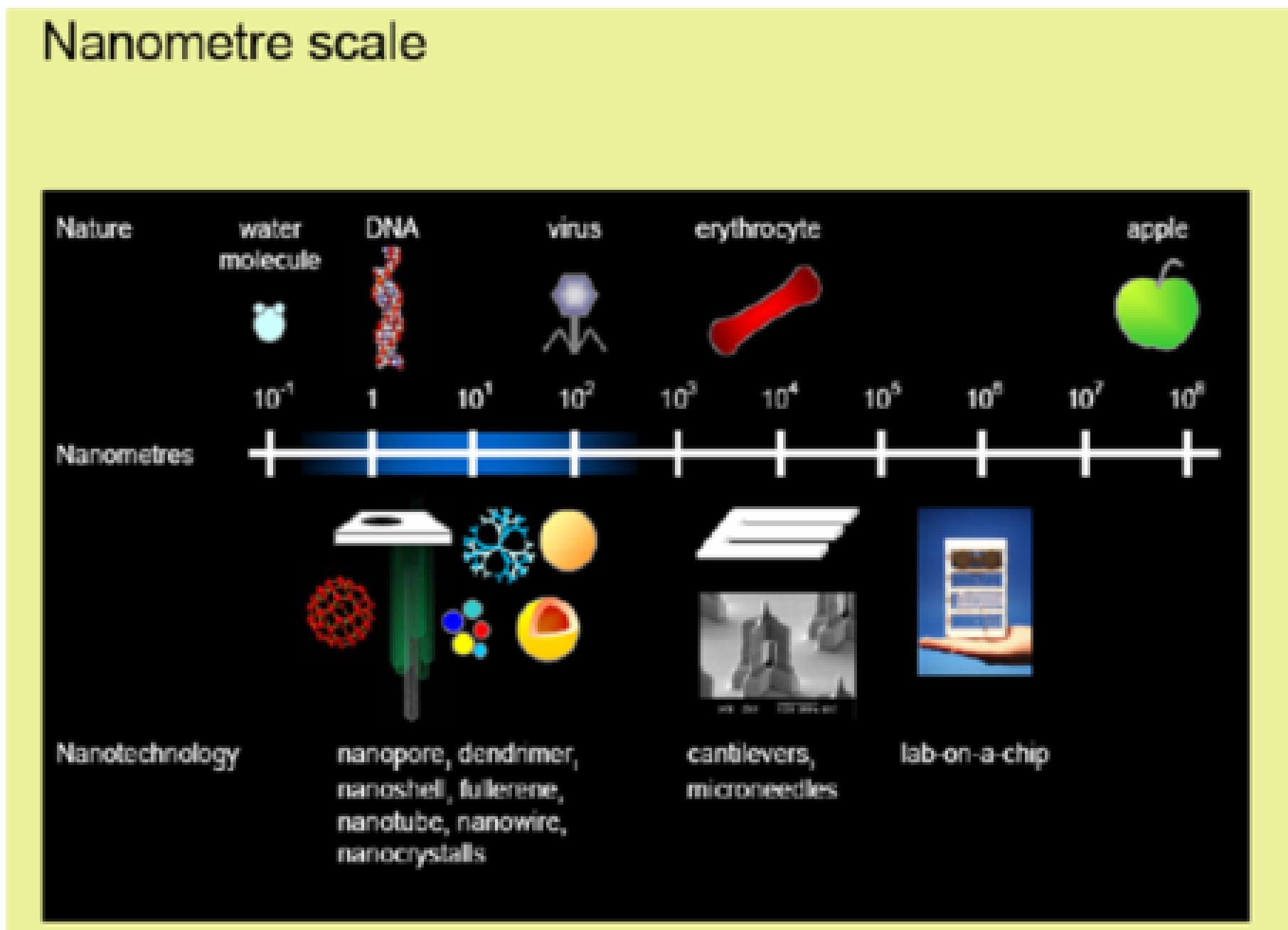

Figure 2. Some manmade and nature made devices and tools on nanometre scale.

Secondly, the physical properties of semiconducting nano crystallites are dominated by the spatial confinements of electronic and vibrational excitations. With decreasing size of crystallites, the gap between Highest Occupied Molecular Orbital (HOMO) & Lowest Unoccupied Molecular Orbital (LUMO) widens [7].

It is also possible to fine tune the crucial *band gap* of ZnO (energy of 3.37eV) by doping with divalent metals ions such as $Mn^{2+}$, $Co^{2+}$ and $Ni^{2+}$ [8].

## 3. Multiplicity of Morphologies of ZnO

Controlled synthesis of nanomaterials with respect to shape and size is crucial for the development of nanotechnologies concerened. Various morphologies of nanostructures such as belts, ribbons, cables, rods, tubes, rings, springs, helices, bows, tetrapods, spirals, needles and films are the *specialty of ZnO nanosystems*. The nano helix and belt are shown in Figure 3a and 3b). These have been synthesized via several techniques, such as, Metal Organic Chemical Vapour Deposition (MOCVD)[9,10], Pulsed Laser Deposition (PLD)[11], Molecular Beam Epitaxy (MBE)[12], Vapor-Liquid-Solid mechanism (VLS) [13], Sol-Gel process[14] and Thermal Annealing method[15].

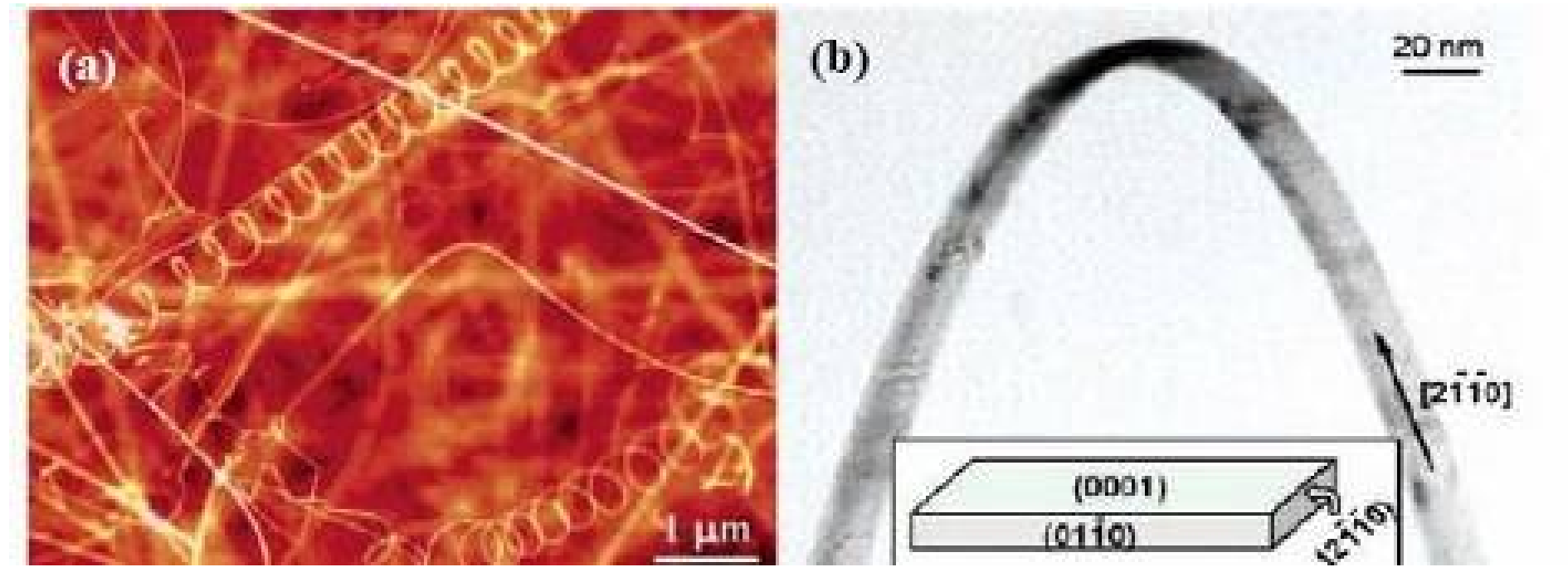

Figure 3. ZnO Nanosystems [16] (a) Helix and (b) Belt

For instance, in a typical MOCVD, a steady increase in temperature of the combustion chamber has resulted in the formation of nanostructures with decreasing diameters [10]. ZnO nano rods with a diameter range 30-200nm were synthesized in a hotwall reactor using Diethyl Zinc transported by high purity Argon. At temp 673K - 773K the size of ZnO was of 1-2 µm. If the temp rises from 773 to 973K, the micro structure was found to evolve into a well aligned shape of around 30nm. If the temperature rises from 1173K to 1273K, the shape of ZnO nano structure changes from nano rods to nano needles or completely random nano wires. By a variation of the controlling parameters, it is possible to get nanocombs & nano sheets as well [10]. In PLD method, the laser power has been utilized as a growth parameter to control the diameter of nanorods by controlling the dimension of 3D nucleation [11]. The MBE method is a catalyst-driven pathway to control the growth of ZnO nanorods. The process is site-specific, as single crystal ZnO nanorod growth is realized via either nucleation on Ag films or deposition on a $SiO_2$, Sapphire or Quartz[12]. The growth of ZnO nanorods via VLS mechanism is based on the bulk diffusion of metal atoms as catalysts [13]. Sol-gel process is known to have the distinct advantage over the other methods because of process simplicity and ease of control of the film composition [14]. The thermal annealing method [15] is relatively the simplest.

## 4. Nanotechnology at DSI

Our research focuses on the fabrication of vertically aligned and uniform ZnO nanorods. The objective is to synthesize perfectly aligned and uniform ZnO nanorods which can be scaled up to practical applications.

In this direction, we have recently prepared ZnO nanorods under specific growth conditions by heating a mixture of zinc acetate and sodium hydroxide pellets in the ratio 1:25 which is favored by hydrothermal oxidation of zinc metal at 120°C for 24 hours.

The Scanning Tunneling Microscope (STM) images of these ZnO nanorods display a random distribution of sizes (Figure 4). The X-ray Diffraction (XRD) pattern recorded over a range of angle 2θ values from 30° to 75° reveals a crystalline wurtzite structure[3] (Figure 5).

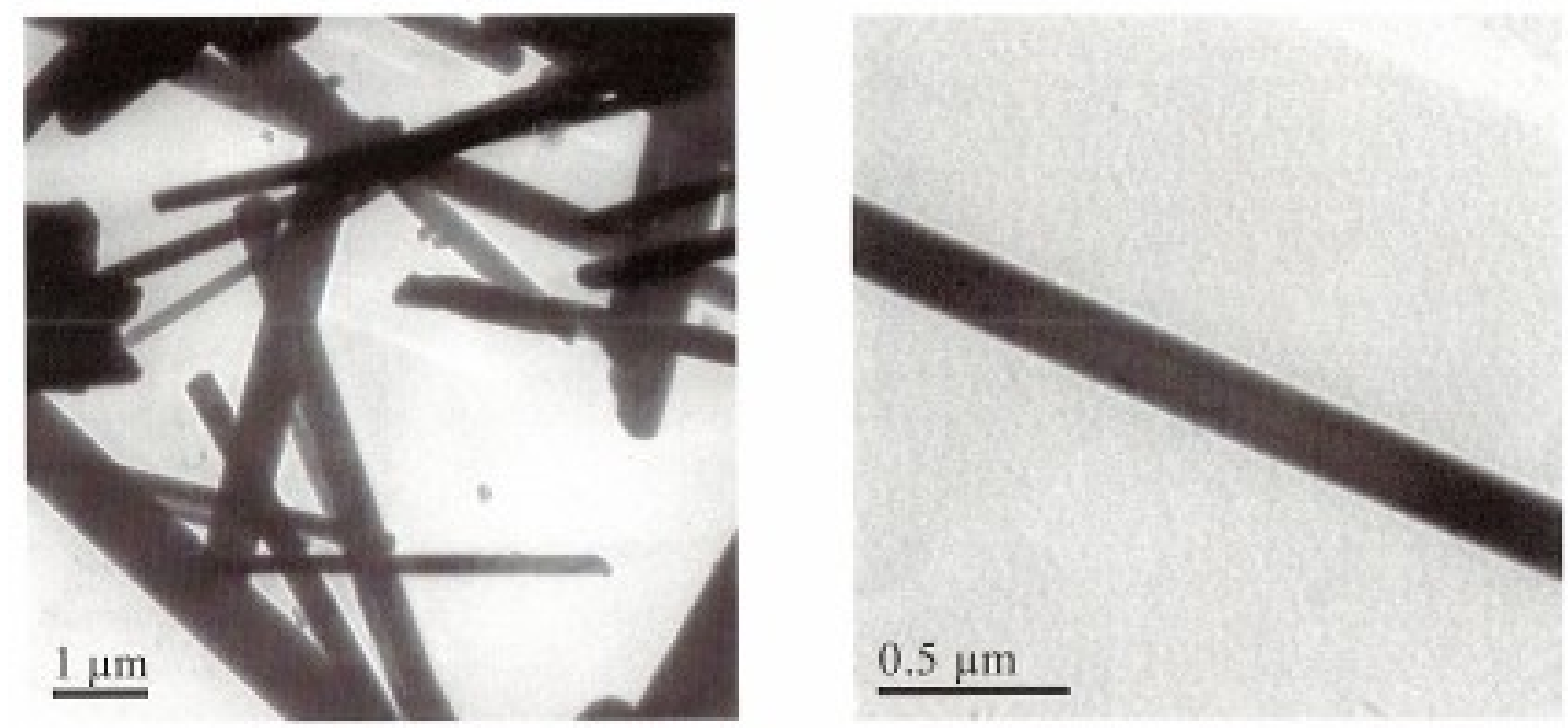

Figure 4. Distribution of ZnO nanorods are shown (size varies from 50 to 150nm)

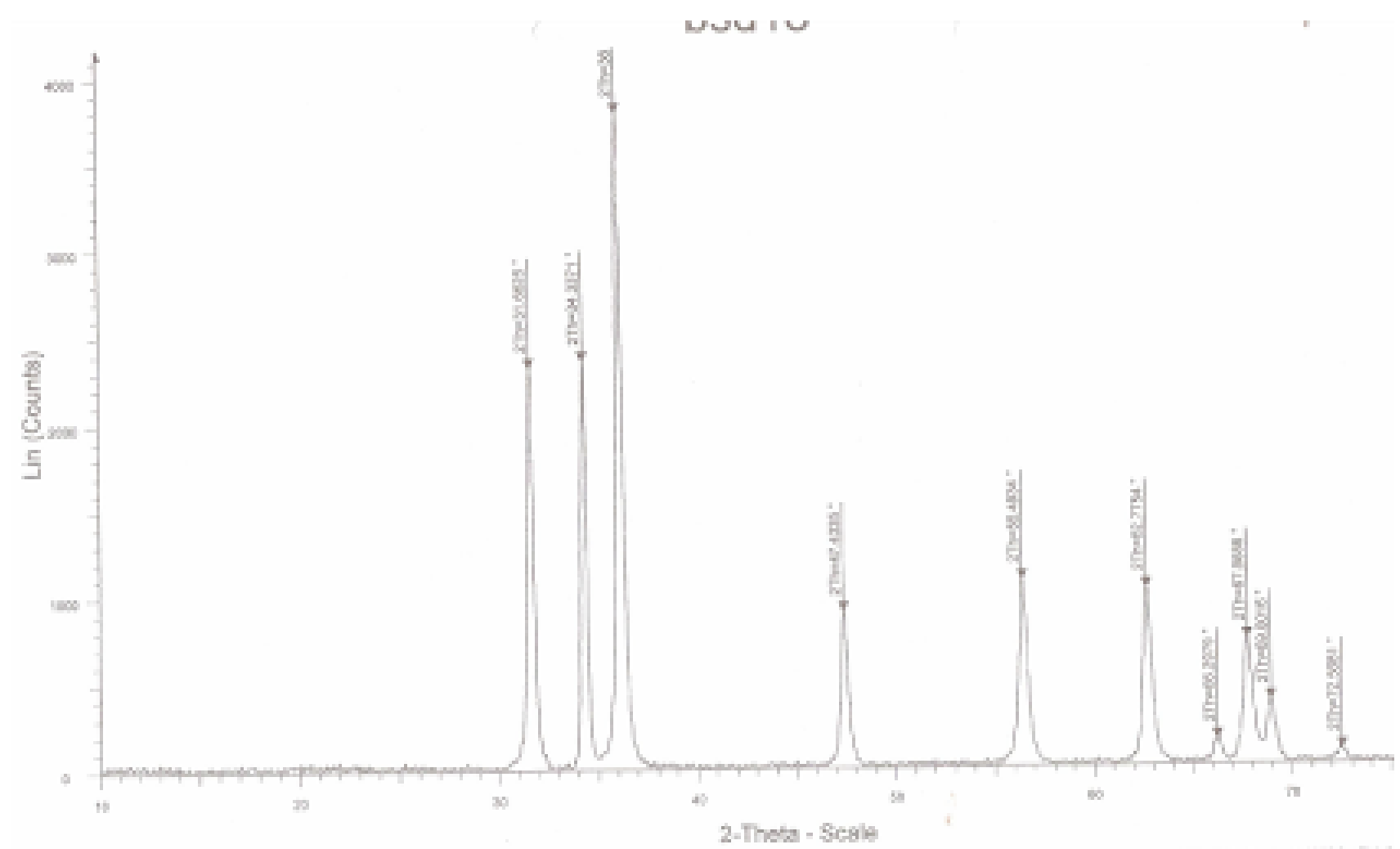

Figure 5. Typical XRD pattern of ZnO nanorods.

Doping with suitable metal ions of ZnO nanoforms to manipulate the electronic properties will be undertaken. Development of indigenous metal catalysts coatings (Au thin film) on substrates (Si, Quartz and Sapphire) suitable for electronic devices is our extended area of research.

**5. Applications of ZnO Nanosystems**

ZnO has been recognized as one of the key materials in Oxide-Electronics [17] enabling the realization of technologies of ultraviolet light emitters, detectors, thin film transistors, spintronics, self organized nanostructures, and so on. ZnO nanowires are extremely sensitive even to tiny forces in the nano- to pico-newton range. This principle is used to make ZnO pressure sensors that can be implanted in the body owing to their biocompatibility [18].

**6. Conclusion**

Nanotechnology is a novel branch of futuristic science and engineering. The synthetic studies of multiple morphologies of ZnO structures constitute the basis for developing versatile applications in the development of new domains.